\documentclass[]{aa}
\pdfoutput=1
\usepackage{graphicx}
\usepackage{amsmath,amsfonts,amssymb,tabu}
\usepackage[para]{threeparttable}
\usepackage{txfonts}
\usepackage[breaklinks,colorlinks,citecolor=blue,pdfa=true]{hyperref}
\usepackage{color}
\usepackage{fixltx2e}
\usepackage{natbib,twoopt}
\usepackage{url}
\usepackage{multirow}
\usepackage{epsf}
\usepackage{epsfig}
\usepackage{longtable}
\usepackage{float}
\usepackage{subfig}
\usepackage{caption}
\newcommand{\GG}[1]{}

\definecolor{mypink1}{RGB}{219, 48, 122}

\usepackage{ifthen}
\usepackage[T1]{fontenc}
\usepackage{lmodern}
\usepackage{ifxetex,ifluatex}
\usepackage{latexsym}
\usepackage{pdfpages}
\usepackage{dblfloatfix}
\usepackage{morefloats}
\usepackage{caption}
\usepackage{lscape}
\usepackage{mathtools}

\bibpunct{(}{)}{;}{a}{}{,} 
\makeatletter
\newcommandtwoopt{\citeads}[3][][]{\href{http://adsabs.harvard.edu/abs/#3}%
{\def\hyper@linkstart##1##2{}%
\let\hyper@linkend\@empty\citealp[#1][#2]{#3}}}
\newcommandtwoopt{\citepads}[3][][]{\href{http://adsabs.harvard.edu/abs/#3}%
{\def\hyper@linkstart##1##2{}%
\let\hyper@linkend\@empty\citep[#1][#2]{#3}}}
\newcommandtwoopt{\citetads}[3][][]{\href{http://adsabs.harvard.edu/abs/#3}%
{\def\hyper@linkstart##1##2{}%
\let\hyper@linkend\@empty\citet[#1][#2]{#3}}}
\newcommandtwoopt{\citeyearads}[3][][]%
{\href{http://adsabs.harvard.edu/abs/#3}
{\def\hyper@linkstart##1##2{}%
\let\hyper@linkend\@empty\citeyear[#1][#2]{#3}}}
\makeatother

\usepackage{graphicx}

\def\simi   {$\sim$\,}

\def \deg         {\text{$^{\circ}$}}
\def \arcmin      {\text{$^\prime$}}

\begin{document}

   \title{Discovery of X-shaped Morphology of the Giant Radio Galaxy GRG 0503-286}

   \titlerunning{Docking of the Tails of the Radio Lobes of GRG 0503-286}

\author {Pratik Dabhade\inst{1}\thanks{E-mail: pratik.dabhade@obspm.fr}
\and Gopal-Krishna\inst{2}
}
  
\institute{$^{1}$Observatoire de Paris, LERMA, Coll\`ege de France, CNRS, PSL University, Sorbonne University, 75014, Paris, France\\
$^{2}$UM-DAE Centre of Excellence in Basic Sciences (CEBS), Vidyanagari, Mumbai - 400098, India\\ 
}

 \date{\today} 
 
 \abstract
{The high surface-brightness sensitivity of the GLEAM survey image of the giant radio galaxy GRG 0503-28 at 70-230 MHz has revealed an inversion-symmetric bending of its two lobes, while maintaining between their bent portions a  \simi200 kpc wide strip-like radio emission gap. This lends the source the  appearance of a mega-sized X-shaped radio galaxy. Identifying the emission gap with the presence of a gaseous layer, probably a WHIM-filled sheet in the cosmic web, we suggest that the layer is the most likely cause of the inversion-symmetric bending of the two radio lobes. Multiple observational manifestations of such gaseous layers are noted. The two lobes of this GRG, known to extend very asymmetrically from the host galaxy, are remarkably symmetric about the emission gap, confirming a curious trend noted earlier for double radio sources of normal dimensions. The anomalous radio spectral gradient reported for the northern lobe of this GRG is not substantiated.}


\maketitle

\section{Introduction} \label{sec:intro}
Starting from their discovery \citep{willis74}, `giant’ radio galaxies (GRGs) have been used as effective probes of the intergalactic medium (IGM). Since their radio lobes typically extend well beyond the virial radius of the host galaxy, they are directly interacting with the warm-hot gaseous phase of the IGM (WHIM; Warm-Hot Intergalactic Medium) which is thought to contain roughly half the baryons in the universe, mostly confined to sheets and filaments delineating the large-scale structure of the universe (e.g., \citealt{Fukugita98,cen99,dave01,shull12}). Until mid-1980s only about a dozen GRGs (size $>$ 1 Mpc\footnote{We adopt a flat cosmology with parameters $\rm \Omega_m$ = 0.27, and a Hubble constant of H$_0$ = 71 km s$^{-1}$ Mpc$^{-1}$.}) were known. Of these, the GRG 0503-286 (MSH05-22) hosted by the luminous elliptical galaxy ESO~422-G028 at $z$=0.03815 and having the radio lobe pair spanning 1.8 Mpc (\simi40\arcmin) was the largest known galaxy in the southern sky. Its discovery\footnote{The authors would like to dedicate this work to late Prof. \href{https://artsandculture.google.com/story/govind-swarup-pioneering-radio-astrophysicist-tata-institute-of-fundamental-research/8wUxlXP1ZillIg?hl=en}{Govind Swarup}, FRS, who took special interest in the discovery of this giant radio galaxy in the southern hemisphere, with the Ooty Synthesis Radio Telescope (OSRT).} was reported independently by two groups (\citet{saripalli86}, hereafter, SGRK86) and \citet{SH86} (SH86). The former paper reported its imaging observations, at 327 MHz with the Ooty Synthesis Radio Telescope (OSRT, \citealt{swarup84osrt}), the precursor to the Giant Metrewave Radio Telescope (GMRT, \citealt{swarup21}), and also at 1.4 GHz and 2.7 GHz using the 100-m Effelsberg Radio Telescope. The moderate luminosity of this GRG (P$_{1.4}$ \simi 1.07 $\times$ 10$^{25}$W~Hz$^{-1}$) places it on the dividing line between FR I and FR II radio sources \citep{FR74}. This is consistent with its lobes lacking a prominent brightness peak and also with its large spectral age of \simi0.3 Gyr (\citet{ravi08}, hereafter SSSH08). Radio polarization imaging of this GRG has been carried out over the frequency range from  about 100 MHz to 5 GHz (SGRK86; \citealt{nvss,Jamrozy05,Riseley18}). The elliptical host galaxy with some young stellar population ($<$10~Myr) in the north-west region, has been classified as a low-excitation-radio-galaxy (LERG; \citealt{Zovaro22}) and low-ionization nuclear emission-line region (LINER; \citealt{veron10}), consistent with the non-detection of [OIII] and H$\beta$ emission lines reported in the discovery papers (SH86; SGRK86). Arguably, this source remains the most assiduously observed GRG, prompted by its highly unusual radio morphology which is marked by the strikingly dissimilar extents and appearances of its two radio lobes, compounded by a large offset and misalignment of the N-lobe from the well defined radio axis of this GRG. SH86 proposed that this GRG may simply be a large `bent-double' radio source seen in projection. However, in SGRK86, a denser gaseous medium inferred from the the higher (projected) concentration of galaxies to the northeast/east of the N-lobe was deemed responsible for both (i) the smaller length of the N-lobe and (ii) for driving this diffuse lobe westward via action of a buoyancy force exerted by the thermal gas associated with the galaxy concentration. As discussed in Sec.~\ref{sec:RD}, this scenario finds support from subsequent optical studies of the large-scale environment of this GRG and also by the distribution of ROSAT detected discrete X-rays sources towards it (\citet{Jamrozy05}, hereafter J05).

\subsection{The galaxy clustering environment and radio lobe asymmetry of GRGs}
The above-mentioned early work on galaxy clustering around the GRG 0503-28 was revisited in a seminal study by SSSH08, by measuring spectroscopic redshifts of over 350 galaxies seen within \simi1\deg of the GRG 0503-28. Its massive elliptical host galaxy was shown to be associated with a small unvirialised group of 4 - 5 galaxies within \simi0.3 Mpc (see, also, 
\citet{Tully15}), from which no extended X-ray emission is detected in the 22 ks {\it ROSAT} pointed observations (J05). SSSH08 detected a \simi100 Mpc long sheet-like filamentary galaxy distribution approaching to within 10-15 Mpc northeast of the host galaxy (see also, \citet{Kalinkov95,jones04}). The VLA maps of this GRG at 1.5 GHz and 4.9 GHz have confirmed the 
very different morphologies of the two lobes and also revealed a \simi90 kpc long straight jet pointing towards the S-lobe (figure 6 of SSSH08), as well as a fainter and shorter counter-jet, both co-linear with the elongated S-lobe. \textit{Thus, the overall radio axis of this GRG is neatly defined (SSSH08)}. It was further argued by these authors that the large displacement of the N-lobe from the radio source axis, towards southwest as well as its misaligned orientation from that axis are plausibly due to the buoyancy force exerted by a denser ambient IGM under the gravitational pull of the afore-mentioned sheet-like galaxy distribution located \simi 10 - 15 Mpc northeast of the N-lobe, with additional push coming from a galactic superwind emanating from a dense galaxy clustering \simi2 Mpc northeast of the host galaxy. In this context it is interesting that as many as six {\it ROSAT} detected discrete X-ray sources without optical counterpart are arrayed along the eastern rim of the N-lobe, in a stark contrast to the S-lobe (J05). Extending the above work, \citet{malarecki15} have reported optical spectroscopy of 9076 galaxies seen in the environments of 19 GRGs, and shown that the shorter lobe lies on the side of larger galaxy number density,  reaffirming the previous studies based on fewer GRGs (SSSH08; \citet{safouris09}; SGRK86). They also found supportive evidence for (i) the GRG lobes to get misaligned from the overall radio axis due to their components being deflected away by galaxy over-densities in the ambient space, and (ii) a tendency for the radio lobes ejected along directions that avoid denser galaxy distributions, to be able to grow to giant sizes. However, a generalisation that higher ambient IGM density inhibits GRG formation has been weakened by the recent detailed study of the Mpc-scale environments of \simi 150 GRGs \citep{tang20,lan21} and, perhaps more directly by the finding that at least 10\% of the GRGs reside in centres of galaxy clusters \citep{PDLOTSS,DabhadeSAGAN20}. This would then appeal to the alternative explanation in which the GRG phenomenon is attributed to a more powerful central engine \citep{gk89,ravi96}, which may conceivably be related to the central engines of GRGs operating at lower Eddington ratios, compared to regular size radio galaxies \citep{DabhadeSAGAN20}.

In Sec.~\ref{sec:RD} we present and discuss new features of the lobes of GRG 0503-286, gleaned from the its low-frequency radio maps of exceptionally high surface-brightness sensitivity, made in the GLEAM survey\footnote{\url{http://gleam-vo.icrar.org/gleam_postage/q/form}} \citep{gleam-walker17} using the Murchison Widefield Array \citep{mwa}. These maps have revealed an X-shaped morphology in this GRG, shedding new light on the underlying mechanism. The main conclusions from this study are summarised in Sec.~\ref{sec:conc}.

\begin{figure}
\centering
\includegraphics[scale=0.25]{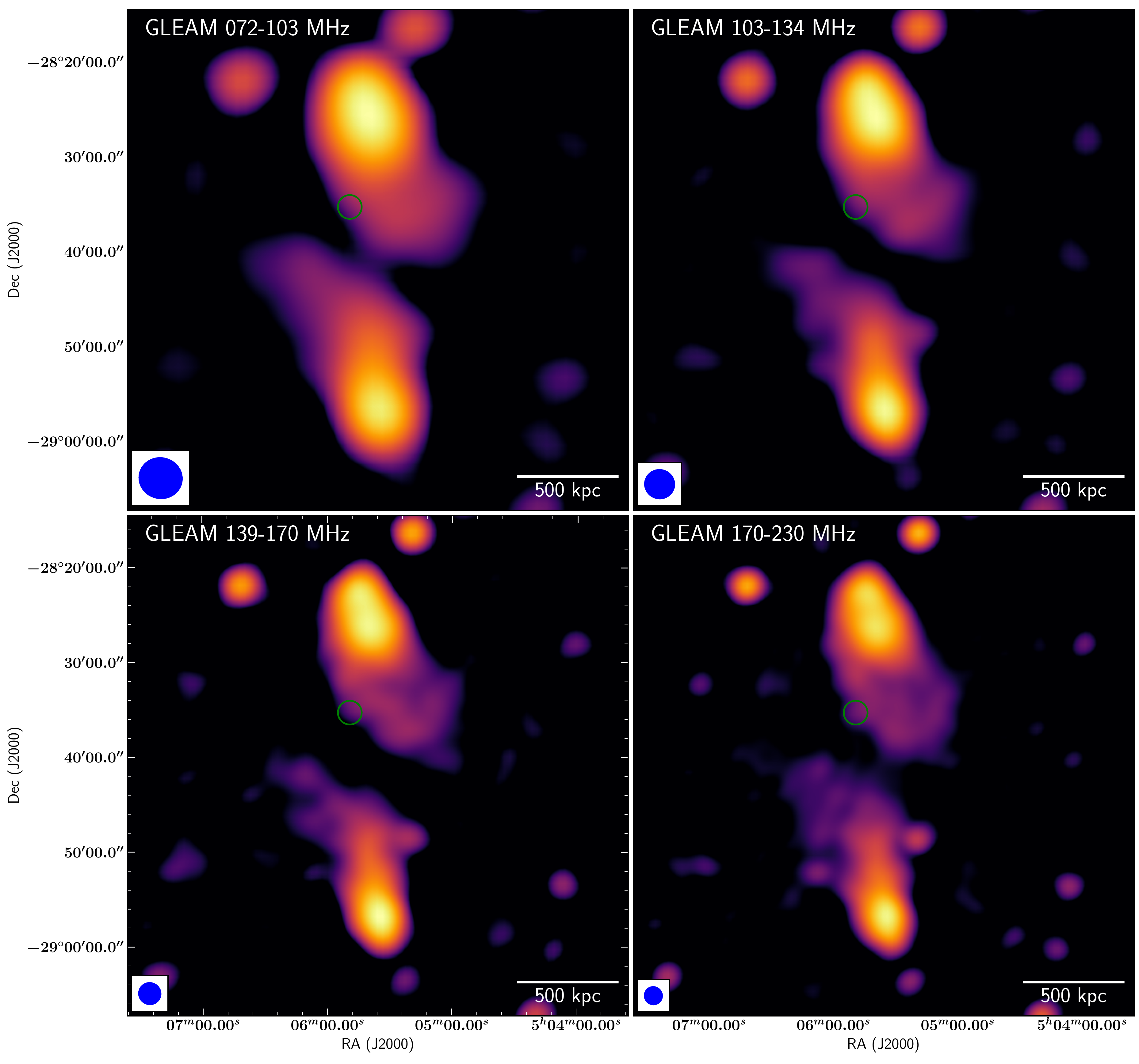}
\caption{The GLEAM radio maps in 4 bands, where the beam is shown in blue circle. The resolutions (FWHM) of the images from 70 to 230 MHz are 4.84\arcmin~$\times$~4.57\arcmin;~77.4$^{\circ}$, 3.43\arcmin~$\times$~3.35\arcmin;~76.2$^{\circ}$, 2.63\arcmin~$\times$~2.58\arcmin;~62.1$^{\circ}$, and 2.19\arcmin~$\times$~2.16\arcmin;~37.3$^{\circ}$.
The location of the host galaxy is marked with a green circle. The radio emission seen in colour is above 3$\sigma$, where $\sigma$ is the rms of each map given in Tab.~\ref
{tab:radioimtab}.}
\label{fig:gims}
\end{figure} 

\vspace{-0.6cm}
\section{Results and discussion}\label{sec:RD}

\subsection{Discovery of inversion-symmetric bending of the lobes}
Fig.~\ref{fig:gims} shows the GLEAM survey maps of the GRG 0503-286 in 4 consecutive frequency bands within 70-230~MHz. These sensitive maps clearly show that approaching towards each other, the two lobes gradually undergo a substantial bending in opposite directions, whereafter they run parallel to each other, maintaining a strip-like radio emission gap between them. The sharp quasi-linear inner edges of the bent lobes (‘spurs’) are more clearly visible in the 3 higher frequency GLEAM maps, due to their superior resolution. This morphology is reminiscent of X-shaped radio galaxies (XRGs; see \citet{xrggk12}, for a review), with the two radio spurs being analogs of the radio `wings’ ( `secondary’ lobes) observed in XRGs. In a currently popular model \citep{leahy84}, the wings form as the hydrodynamic backflow of synchrotron plasma in the twin radio lobes is deflected 
sideways  (see below). Although this model is challenged by XRGs in which the wings are distinctly longer than the primary lobes (see, e.g., \citealt{lp91}), a possible explanation for this can be found in \citet{Capetti02,kraft05,Hodges11}.

By far the most spectacular example of an X-shaped GRG is seen in the recent MeerKAT map of the PKS~2014-55 (size \simi1.57~Mpc), where the hydrodynamic backflow in each lobe appears to undergo a sharp deflection (almost a reflection!), imparting the source the appearance of a `double boomerang' \citep{cotton20}. Although, there is no trace of hot/warm spots in the lobes, this FR I morphology of both its `primary' lobes may still be reconciled with the backflow model by postulating that the primary lobes had an FR II past \citep{saripalli09}. Whilst the possibility of morphological transition from FR II to FR I, due to an interplay between the jet power and the density/temperature of the external medium, has been deemed feasible since long \citep{GKW88NAT,GK91,Bicknell95,gk01F}, the sustenance of a well-collimated backflow after the demise of the Mach disk (hot spot) remains to be demonstrated.

The hydrodynamic backflow model invoking a large halo of coronal gas around the host elliptical galaxy \citep{leahy84} gets stretched when applied to the present GRG 0503-28. Not only is the combined lateral span of its two lobes, sought to be deflected, is huge (\simi0.9~Mpc), necessitating a cluster-size halo of hot gas, but even the region which the two lobes approach before getting deflected in opposite directions, is devoid of a conspicuous galaxy. The only massive galaxy seen between the two lobes is the parent galaxy of the GRG (as marked in Fig.~\ref{fig:gims}) and it is situated far away (\simi350 kpc) from the point of inversion symmetry of the deflected lobe pair. The galaxy is, in fact, located at the boundary of the strip-like radio emission gap (Sec.~\ref{sec:gap}). Several such intriguing examples are discussed in \citet{GK09}. Note also that post-bending, the sharp quasi-linear inner edges of the two lobes run almost parallel to each other for \simi 700 
kpc, end to end, separated by the emission gap (Fig.~\ref{fig:gims}). All this suggests that the backflows in the two radio lobes are being docked by a fat pancake type gaseous layer, possibly a segment of a sheet of the cosmic web permeated by WHIM (Sec.~\ref{sec:2.2}). This possibility has been proposed in earlier studies, together with the stipulation that the putative gaseous layer should be structure of a quasi-permanent nature which does not get disrupted due to any motion of the host galaxy, away from the symmetry region  \citep{GK00,GK09}. In the present case of GRG 0503-286, such a movement of the host galaxy seems implausible, anyway, given its location right on the radio axis so robustly defined by the alignment of the \textit{straight} radio jet, counter-jet and the elongated S-lobe (Figs. 4 and 6 of SSSH08). This goes to suggest that the origin of the gaseous layer invoked to explain the radio emission gap is probably not linked to the host galaxy itself, rather the layer has existed independently of the galaxy and could well be a WHIM-permeated sheet of the cosmic web (Sec.~\ref{sec:2.2}).  

\begin{table}[htbp]

\begin{minipage}{90mm}
\captionsetup{width=8cm}
\caption{Flux densities and rms of the  GRG 0503-286 from the 4 GLEAM maps. The notations Flux$_{\rm T}$, Flux$_{\rm N}$, and Flux$_{\rm S}$ refer to flux densities of the entire source, northern lobe, and southern lobe, respectively. }\label{tab:radioimtab}

\begin{tabular}{lcccc}
\hline
 Frequency & RMS  & Flux$_{\rm T}$ &Flux$_{\rm N}$ & Flux$_{\rm S}$ \\ 
  (MHz)    & (Jy) & (Jy) & (Jy) & (Jy) \\ 

\hline
72-103 & 0.030  & 20.8$\pm$1.1 & 11.9$\pm$1.2 & 8.9$\pm$0.9\\
103-134 & 0.020 & 16.7$\pm$0.8 & 9.6$\pm$0.9 & 7.1$\pm$0.7\\
139-170 & 0.012 &12.9$\pm$0.7  & 7.5$\pm$0.7 & 5.4$\pm$0.5\\
170-230 & 0.008 & 12.1$\pm$0.6 & 6.9$\pm$0.7 & 5.2$\pm$0.5 \\

\hline
\end{tabular}
\end{minipage}
\end{table}

\subsection{Does the sharp radio emission gap define the symmetry plane in radio galaxies?}\label{sec:2.2}
Well over a dozen FR II radio galaxies have been reported to exhibit strip-like regions of depressed radio emission (typical width \simi30~kpc, but with a wide range) separating the two radio lobes and usually running almost orthogonally to the axis defined by them \citep{GK00,gkw07,GK09}. 
Since the quasi-linear sharp edges seen between the lobes in these double radio sources would only be observable when the radio axis is oriented near the plane of the sky (in order that the twin radio lobes do not appear partially overlapping), such emission gaps and the proposed gaseous `superdisks/fat-pancakes/cosmic-sheets' causing them, are probably associated with many more radio galaxies than found so far. As summarised in these papers, the well known phenomena for which the proposed fat-pancakes may provide viable alternative explanation, 
include: (i) the radio lobe depolarisation asymmetry, popularly known as Laing-Garrington effect \citep{Garrington88,laing88} (see, \citealt{gknath97}); (ii) the correlated radio-optical asymmetry of double radio sources \citep{McCarthy91} (see, \citealt{GKW96}); (iii) the occurrence of absorption dips in the Ly-$\alpha$ emission profiles of high-$z$ radio galaxies with a total radio extent of up to \simi50~kpc \citep{vanOjik97,Binette06} (see \citealt{GK00}); and (iv) the apparent asymmetry of the extended Ly-$\alpha$ emission associated with 
the lobes of high-$z$ radio galaxies (see \citealt{GK00,GK04}). It was also argued that the proposed fat pancakes may be instrumental in causing metre-wavelength flux variability via `superluminal refractive scintillations' \citep{GK91} see, also \citep{Campbell94,Ferrara01}.

Further, the observed strip-like radio emission gaps in radio galaxies  bear a close relevance to the question `what docks the tails of radio source components in double radio sources?', raised by \citet{Jenkins76} who concluded that the answer to this cannot be synchrotron losses in the lobes. In fact, all these characteristics are connected to the more general and long standing issue of asymmetries in double radio sources (reviewed, e.g., in \citet{GK04}). Here, it is interesting to recall the curious, statistically significant trend that the two lobes in FR II radio galaxies are found to extend more symmetrically with respect to the radio emission gap than they do about the host galaxy \citep{GK00,GK09}. While the cause of this behaviour remains obscure, it may be noted that the present GRG is in full conformity to this trend. Even though, its lobe-length ratio is exceptionally large (1 : 1.6, SSSH08) among GRGs \citep{DabhadeSAGAN20}, when measured relative to the host galaxy, the lobe-length asymmetry virtually disappears when referenced to the mid-plane of the emission gap (Fig.~\ref{fig:gims}).

\begin{figure}
\centering
\includegraphics[scale=0.32]{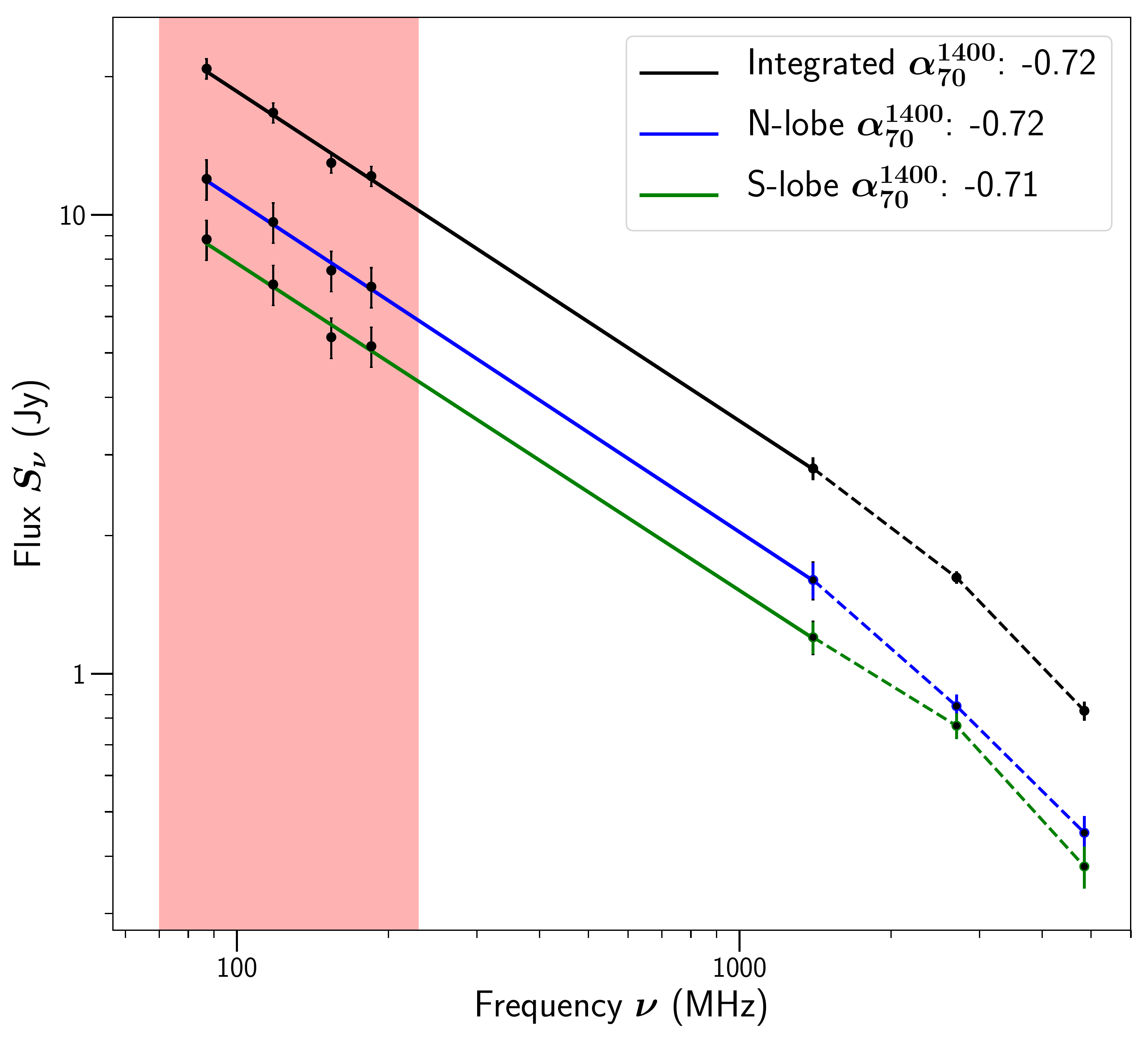}
\caption{The plot shows spectral index fit for flux densities of the GRG using GLEAM (Tab.~\ref
{tab:radioimtab}), Effelsberg (1.4 GHz and 2.7 GHz from SGRK86), and PMN survey at 5~GHz (SSSH08). The frequency range covered by GLEAM is shown by red shaded area. }
\label{fig:SIFIT}
\end{figure}

\begin{figure*}
\centering
\includegraphics[scale=0.11]{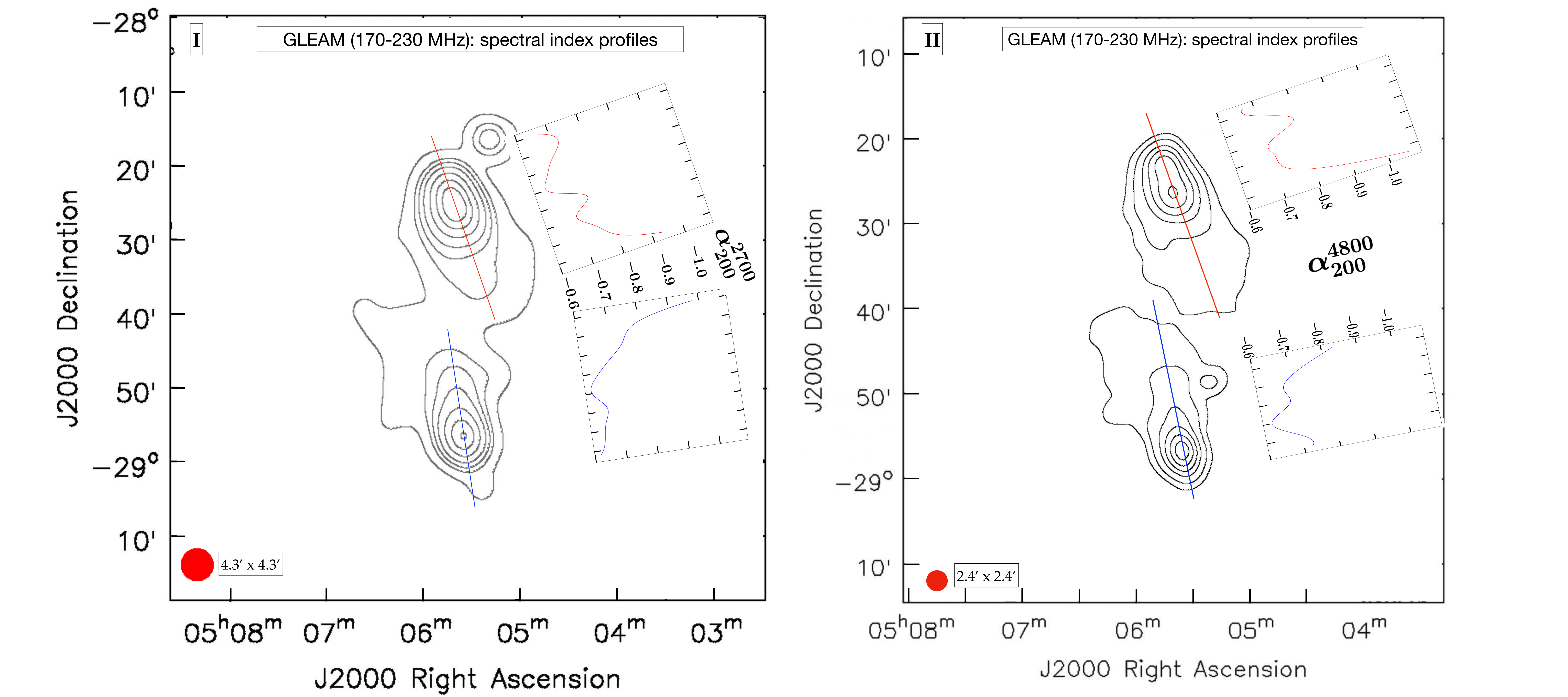}
\caption{170-230 MHz GLEAM contour maps with 4.3\arcmin (I; left side) and 2.4\arcmin (II; right side) resolutions showing spectral index ($\alpha_{200}^{2700}$ and $\alpha_{200}^{4800}$) profiles along the axes of the two lobes (insets). For subplots a and b in I and II, the tick marks along the vertical axis are separated by \simi2\arcmin.}
\label{fig:SIGRAD}
\end{figure*}

\subsection{What fills the radio emission gaps?}\label{sec:gap}
The issue of the content of the gaseous layer, inferred using the radio emission gap as the signpost, was recently discussed by \citet{anand19} who also presented the negative results of their VLA search for HI signal from the conspicuous emission gaps seen in 4 nearby radio galaxies (z $=$ 0.03 to 0.09).
Early hints that such gaseous layers (fat pancakes) are dusty, emerged from the observed stark asymmetry of the extended Ly-$\alpha$ emission associated with the twin-lobes of high-$z$ radio galaxies \citep{GK00}, and also from their correlated radio-optical asymmetry \citep{GKW96} (Sec~\ref{sec:2.2}). A potentially interesting additional hint about the content of the putative gaseous layer in the GRG 0503-286 comes from the {\it ROSAT} X-ray image of this field (J05) which shows a concentration of about 10 discrete soft X-ray sources in the region of the putative layer (emission gap) between the two lobes. The optical identifications of these soft X-ray sources can be regarded as tentative, given the substantial uncertainty of the {\it ROSAT} positions. Thus, while it remains to be demonstrated if any of these X-ray sources are actually embedded within the putative gaseous layer, the localised enhancement in their surface density in this region 
indicates that some of them may even turn out to be peaks within an extended faint X-ray emission associated with the layer. This preliminary indication calls for a sensitive X-ray imaging of this field.

\subsection{Does the GRG 0503-286 show an anomalous spectral index gradient ?}

Fig.~\ref{fig:SIFIT} shows the radio spectra of the entire source and its two lobes, based on the flux densities listed in Tab.~\ref{tab:radioimtab} and referenced in the caption. The integrated spectrum can be fitted with two straight lines, with a slope $\alpha$ = -0.72$\pm$0.02 up to \simi2.7~GHz and -0.98$\pm$0.06 thereafter, until \simi5~GHz. A similar pattern can be discerned for the two lobes. The compact radio core associated with the host galaxy and the twin-jets, are only detected clearly in the VLA observations (SSSH08) and together contribute a 
negligibly small fraction of the total emission (\simi0.35\% at 5~GHz). 

SSSH08 have presented spectral index distribution along the two lobes, based on their maps at 843 MHz (MOST) and 1520 MHz (VLA, CnB array), at a common resolution of \simi1\arcmin. They find an unusual pattern for the N-lobe. While, its spectrum steepens towards the leading rim, it shows a flattening towards the host galaxy, in contrast to the pattern established for 
the lobes of edge-brightened double radio sources, and indeed also found by SSSH08 for the S-lobe of this GRG. This spectral contrast between the two lobes would add a further new dimension to the lobe asymmetry in this GRG. 
To examine this, we have compared the 4.8~GHz (FWHM = 2.4\arcmin ; J05) and 2.7~GHz (FWHM = 4.3\arcmin ; SGRK86) Effelsberg maps with the 170 - 230 MHz GLEAM map
(FWHM =2.19\arcmin~$\times$~2.16\arcmin; Fig.~\ref{fig:gims}) after smoothing the GLEAM map to the resolution of the respective Effelsberg maps, as shown in Fig.~\ref{fig:SIGRAD}. 
The enormous frequency range (1 : 24) covered by the maps lends a high degree of spectral precision (a 10\% error in the ratio of flux densities at the two frequencies corresponds to a change of just 0.03 in the spectral index). Fig.~\ref{fig:SIGRAD} shows the spectral index profiles along the axes of the northern and southern lobes marked with red and blue lines. A small but significant steepening towards the leading edge is observed for both lobes. However, we find no indication that in the N-lobe, $\alpha_{200}^{4800}$ and  $\alpha_{200}^{2700}$ flattens 
towards the host galaxy; the opposite is seen, in conformity with the spectral gradient reported in SGRK86. The origin of the discrepancy with the finding of SSSH08 remains to be understood. Indeed, the steepening trend of the spectral profiles of both lobes (Fig.~\ref{fig:SIGRAD}), as inferred from the 3 most suitable radio maps available, is in accord with the expectation of a steeper spectrum for the wings.

\vspace{-0.6cm}
\section{Conclusions} \label{sec:conc}
The recent GLEAM survey maps of the giant radio galaxy GRG 0503-28, covering the 70 - 230 MHz band with a very high surface-brightness sensitivity, has enabled us to revisit some of the reported intriguing aspects of its exceptional lobe asymmetry. These new maps unveil an X-shaped radio morphology of this GRG, by revealing an inversion-symmetric bending of its radio lobes, while maintaining a strip-like emission gap between their bent portions. If such a bending pattern were to arise from deflection of the backflow of the two lobes, in accord with a currently popular model, the only candidate massive galaxy that could have caused this is the host galaxy of the GRG. However, it is situated \simi350~kpc from the point of inversion symmetry of the bent radio lobes, near the northern edge of the emission gap and right on the robustly defined radio axis of this GRG. This strongly disfavours the possibility of the galaxy having migrated from the point of inversion-symmetry to its present faraway location. We are led to posit that both the inversion-symmetric lobe bending and the strip-like radio emission gap between the two bent lobes have a common cause, probably a gaseous layer which might be a WHIM permeated sheet of the cosmic web. Various observational evidences and results in support of this proposal are summarised, including the one stemming from the existing {\it ROSAT} X-ray observations of the present GRG. Further, it is noted that the GRG~0503-28, known for the exceptionally unequal extents of its two lobes, in fact extends symmetrically about the radio emission gap. This agrees with the previously reported trend for double radio sources of normal dimensions. Lastly, the reported anomalous radio spectral gradient in the northern lobe is not substantiated in this study.

\section*{Acknowledgements}
GK acknowledges a Senior Scientist fellowship of the Indian National Science Academy. We acknowledge that this work has made use of \textsc{aplpy} \citep{apl}.

\bibliographystyle{aa} 
\bibliography{GRG0503-286_GK.bib}

\begin{thebibliography}{57}
\expandafter\ifx\csname natexlab\endcsname\relax\def\natexlab#1{#1}\fi

\bibitem[{{Anand} {et~al.}(2019){Anand}, {Roy}, \& {Gopal-Krishna}}]{anand19}
{Anand}, A., {Roy}, N., \& {Gopal-Krishna}. 2019, Research in Astronomy and
  Astrophysics, 19, 083

\bibitem[{{Bicknell}(1995)}]{Bicknell95}
{Bicknell}, G.~V. 1995, \apjs, 101, 29

\bibitem[{{Binette} {et~al.}(2006){Binette}, {Wilman}, {Villar-Mart{\'\i}n},
  {Fosbury}, {Jarvis}, \& {R{\"o}ttgering}}]{Binette06}
{Binette}, L., {Wilman}, R.~J., {Villar-Mart{\'\i}n}, M., {et~al.} 2006, \aap,
  459, 31

\bibitem[{{Campbell-Wilson} \& {Hunstead}(1994)}]{Campbell94}
{Campbell-Wilson}, D. \& {Hunstead}, R.~W. 1994, \pasa, 11, 33

\bibitem[{{Capetti} {et~al.}(2002){Capetti}, {Zamfir}, {Rossi}, {Bodo},
  {Zanni}, \& {Massaglia}}]{Capetti02}
{Capetti}, A., {Zamfir}, S., {Rossi}, P., {et~al.} 2002, \aap, 394, 39

\bibitem[{{Cen} \& {Ostriker}(1999)}]{cen99}
{Cen}, R. \& {Ostriker}, J.~P. 1999, \apj, 514, 1

\bibitem[{{Condon} {et~al.}(1998){Condon}, {Cotton}, {Greisen}, {Yin},
  {Perley}, {Taylor}, \& {Broderick}}]{nvss}
{Condon}, J.~J., {Cotton}, W.~D., {Greisen}, E.~W., {et~al.} 1998, \aj, 115,
  1693

\bibitem[{{Cotton} {et~al.}(2020){Cotton}, {Thorat}, {Condon}, {Frank},
  {J{\'o}zsa}, {White}, {Deane}, {Oozeer}, {Atemkeng}, {Bester}, {Fanaroff},
  {Kupa}, {Smirnov}, {Mauch}, {Krishnan}, \& {Camilo}}]{cotton20}
{Cotton}, W.~D., {Thorat}, K., {Condon}, J.~J., {et~al.} 2020, \mnras, 495,
  1271

\bibitem[{{Dabhade} {et~al.}(2020{\natexlab{a}}){Dabhade}, {Mahato}, {Bagchi},
  {Saikia}, {Combes}, {Sankhyayan}, {R{\"o}ttgering}, {Ho}, {Gaikwad},
  {Raychaudhury}, {Vaidya}, \& {Guiderdoni}}]{DabhadeSAGAN20}
{Dabhade}, P., {Mahato}, M., {Bagchi}, J., {et~al.} 2020{\natexlab{a}}, \aap,
  642, A153

\bibitem[{{Dabhade} {et~al.}(2020{\natexlab{b}}){Dabhade}, {R{\"o}ttgering},
  {Bagchi}, {Shimwell}, {Hardcastle}, {Sankhyayan}, {Morganti}, {Jamrozy},
  {Shulevski}, \& {Duncan}}]{PDLOTSS}
{Dabhade}, P., {R{\"o}ttgering}, H.~J.~A., {Bagchi}, J., {et~al.}
  2020{\natexlab{b}}, \aap, 635, A5

\bibitem[{{Dav{\'e}} {et~al.}(2001){Dav{\'e}}, {Cen}, {Ostriker}, {Bryan},
  {Hernquist}, {Katz}, {Weinberg}, {Norman}, \& {O'Shea}}]{dave01}
{Dav{\'e}}, R., {Cen}, R., {Ostriker}, J.~P., {et~al.} 2001, \apj, 552, 473

\bibitem[{{Fanaroff} \& {Riley}(1974)}]{FR74}
{Fanaroff}, B.~L. \& {Riley}, J.~M. 1974, \mnras, 167, 31P

\bibitem[{{Ferrara} \& {Perna}(2001)}]{Ferrara01}
{Ferrara}, A. \& {Perna}, R. 2001, \mnras, 325, 1643

\bibitem[{{Fukugita} {et~al.}(1998){Fukugita}, {Hogan}, \&
  {Peebles}}]{Fukugita98}
{Fukugita}, M., {Hogan}, C.~J., \& {Peebles}, P.~J.~E. 1998, \apj, 503, 518

\bibitem[{{Garrington} {et~al.}(1988){Garrington}, {Leahy}, {Conway}, \&
  {Laing}}]{Garrington88}
{Garrington}, S.~T., {Leahy}, J.~P., {Conway}, R.~G., \& {Laing}, R.~A. 1988,
  \nat, 331, 147

\bibitem[{{Gopal-Krishna}(1991)}]{GK91}
{Gopal-Krishna}. 1991, \aap, 248, 415

\bibitem[{{Gopal-Krishna} {et~al.}(2012){Gopal-Krishna}, {Biermann}, {Gergely},
  \& {Wiita}}]{xrggk12}
{Gopal-Krishna}, {Biermann}, P.~L., {Gergely}, L.~{\'A}., \& {Wiita}, P.~J.
  2012, Research in Astronomy and Astrophysics, 12, 127

\bibitem[{{Gopal-Krishna} \& {Nath}(1997)}]{gknath97}
{Gopal-Krishna} \& {Nath}, B.~B. 1997, \aap, 326, 45

\bibitem[{{Gopal-Krishna} \& {Wiita}(1988)}]{GKW88NAT}
{Gopal-Krishna} \& {Wiita}, P.~J. 1988, \nat, 333, 49

\bibitem[{{Gopal-Krishna} \& {Wiita}(1996)}]{GKW96}
{Gopal-Krishna} \& {Wiita}, P.~J. 1996, \apj, 467, 191

\bibitem[{{Gopal-Krishna} \& {Wiita}(2000)}]{GK00}
{Gopal-Krishna} \& {Wiita}, P.~J. 2000, \apj, 529, 189

\bibitem[{{Gopal-Krishna} \& {Wiita}(2001)}]{gk01F}
{Gopal-Krishna} \& {Wiita}, P.~J. 2001, \aap, 373, 100

\bibitem[{{Gopal-Krishna} \& {Wiita}(2004)}]{GK04}
{Gopal-Krishna} \& {Wiita}, P.~J. 2004, arXiv e-prints, 0409761

\bibitem[{{Gopal-Krishna} \& {Wiita}(2009)}]{GK09}
{Gopal-Krishna} \& {Wiita}, P.~J. 2009, \na, 14, 51

\bibitem[{{Gopal-Krishna} {et~al.}(2007){Gopal-Krishna}, {Wiita}, \&
  {Joshi}}]{gkw07}
{Gopal-Krishna}, {Wiita}, P.~J., \& {Joshi}, S. 2007, \mnras, 380, 703

\bibitem[{{Gopal-Krishna} {et~al.}(1989){Gopal-Krishna}, {Wiita}, \&
  {Saripalli}}]{gk89}
{Gopal-Krishna}, {Wiita}, P.~J., \& {Saripalli}, L. 1989, \mnras, 239, 173

\bibitem[{{Hodges-Kluck} \& {Reynolds}(2011)}]{Hodges11}
{Hodges-Kluck}, E.~J. \& {Reynolds}, C.~S. 2011, \apj, 733, 58

\bibitem[{{Hurley-Walker} {et~al.}(2017){Hurley-Walker}, {Callingham},
  {Hancock}, {Franzen}, {Hindson}, {Kapi{\'n}ska}, {Morgan}, {Offringa},
  {Wayth}, {Wu}, {Zheng}, {Murphy}, {Bell}, {Dwarakanath}, {For}, {Gaensler},
  {Johnston-Hollitt}, {Lenc}, {Procopio}, {Staveley-Smith}, {Ekers}, {Bowman},
  {Briggs}, {Cappallo}, {Deshpande}, {Greenhill}, {Hazelton}, {Kaplan},
  {Lonsdale}, {McWhirter}, {Mitchell}, {Morales}, {Morgan}, {Oberoi}, {Ord},
  {Prabu}, {Shankar}, {Srivani}, {Subrahmanyan}, {Tingay}, {Webster},
  {Williams}, \& {Williams}}]{gleam-walker17}
{Hurley-Walker}, N., {Callingham}, J.~R., {Hancock}, P.~J., {et~al.} 2017,
  \mnras, 464, 1146

\bibitem[{{Jamrozy} {et~al.}(2005){Jamrozy}, {Kerp}, {Klein}, {Mack}, \&
  {Saripalli}}]{Jamrozy05}
{Jamrozy}, M., {Kerp}, J., {Klein}, U., {Mack}, K.~H., \& {Saripalli}, L. 2005,
  Baltic Astronomy, 14, 399

\bibitem[{{Jenkins} \& {Scheuer}(1976)}]{Jenkins76}
{Jenkins}, C.~L. \& {Scheuer}, P.~A.~G. 1976, \mnras, 174, 327

\bibitem[{{Jones} {et~al.}(2004){Jones}, {Saunders}, {Colless}, {Read},
  {Parker}, {Watson}, {Campbell}, {Burkey}, {Mauch}, {Moore}, {Hartley},
  {Cass}, {James}, {Russell}, {Fiegert}, {Dawe}, {Huchra}, {Jarrett}, {Lahav},
  {Lucey}, {Mamon}, {Proust}, {Sadler}, \& {Wakamatsu}}]{jones04}
{Jones}, D.~H., {Saunders}, W., {Colless}, M., {et~al.} 2004, \mnras, 355, 747

\bibitem[{{Kalinkov} \& {Kuneva}(1995)}]{Kalinkov95}
{Kalinkov}, M. \& {Kuneva}, I. 1995, \aaps, 113, 451

\bibitem[{{Kraft} {et~al.}(2005){Kraft}, {Hardcastle}, {Worrall}, \&
  {Murray}}]{kraft05}
{Kraft}, R.~P., {Hardcastle}, M.~J., {Worrall}, D.~M., \& {Murray}, S.~S. 2005,
  \apj, 622, 149

\bibitem[{{Laing}(1988)}]{laing88}
{Laing}, R.~A. 1988, \nat, 331, 149

\bibitem[{{Lan} \& {Xavier Prochaska}(2021)}]{lan21}
{Lan}, T.-W. \& {Xavier Prochaska}, J. 2021, \mnras, 502, 5104

\bibitem[{{Leahy} \& {Perley}(1991)}]{lp91}
{Leahy}, J.~P. \& {Perley}, R.~A. 1991, \aj, 102, 537

\bibitem[{{Leahy} \& {Williams}(1984)}]{leahy84}
{Leahy}, J.~P. \& {Williams}, A.~G. 1984, \mnras, 210, 929

\bibitem[{{Malarecki} {et~al.}(2015){Malarecki}, {Jones}, {Saripalli},
  {Staveley-Smith}, \& {Subrahmanyan}}]{malarecki15}
{Malarecki}, J.~M., {Jones}, D.~H., {Saripalli}, L., {Staveley-Smith}, L., \&
  {Subrahmanyan}, R. 2015, \mnras, 449, 955

\bibitem[{{McCarthy} {et~al.}(1991){McCarthy}, {van Breugel}, \&
  {Kapahi}}]{McCarthy91}
{McCarthy}, P.~J., {van Breugel}, W., \& {Kapahi}, V.~K. 1991, \apj, 371, 478

\bibitem[{{Riseley} {et~al.}(2018){Riseley}, {Lenc}, {Van Eck}, {Heald},
  {Gaensler}, {Anderson}, {Hancock}, {Hurley-Walker}, {Sridhar}, \&
  {White}}]{Riseley18}
{Riseley}, C.~J., {Lenc}, E., {Van Eck}, C.~L., {et~al.} 2018, \pasa, 35, e043

\bibitem[{{Robitaille} \& {Bressert}(2012)}]{apl}
{Robitaille}, T. \& {Bressert}, E. 2012, {APLpy: Astronomical Plotting Library
  in Python}

\bibitem[{{Safouris} {et~al.}(2009){Safouris}, {Subrahmanyan}, {Bicknell}, \&
  {Saripalli}}]{safouris09}
{Safouris}, V., {Subrahmanyan}, R., {Bicknell}, G.~V., \& {Saripalli}, L. 2009,
  \mnras, 393, 2

\bibitem[{{(Saripalli} {et~al.}(1986){(Saripalli}, {Gopal-Krishna}, {Reich}, \&
  {Kuehr}}]{saripalli86}
{(Saripalli}, L., {Gopal-Krishna}, {Reich}, W., \& {Kuehr}, H. 1986, \aap, 170,
  20

\bibitem[{{Saripalli} \& {Subrahmanyan}(2009)}]{saripalli09}
{Saripalli}, L. \& {Subrahmanyan}, R. 2009, \apj, 695, 156

\bibitem[{{Shull} {et~al.}(2012){Shull}, {Smith}, \& {Danforth}}]{shull12}
{Shull}, J.~M., {Smith}, B.~D., \& {Danforth}, C.~W. 2012, \apj, 759, 23

\bibitem[{{Subrahmanya} \& {Hunstead}(1986)}]{SH86}
{Subrahmanya}, C.~R. \& {Hunstead}, R.~W. 1986, \aap, 170, 27

\bibitem[{{Subrahmanyan} {et~al.}(1996){Subrahmanyan}, {Saripalli}, \&
  {Hunstead}}]{ravi96}
{Subrahmanyan}, R., {Saripalli}, L., \& {Hunstead}, R.~W. 1996, \mnras, 279,
  257

\bibitem[{{(Subrahmanyan} {et~al.}(2008){(Subrahmanyan}, {Saripalli},
  {Safouris}, \& {Hunstead}}]{ravi08}
{(Subrahmanyan}, R., {Saripalli}, L., {Safouris}, V., \& {Hunstead}, R.~W.
  2008, \apj, 677, 63

\bibitem[{{Swarup}(1984)}]{swarup84osrt}
{Swarup}, G. 1984, Journal of Astrophysics and Astronomy, 5, 139

\bibitem[{{Swarup}(2021)}]{swarup21}
{Swarup}, G. 2021, \araa, 59

\bibitem[{{Tang} {et~al.}(2020){Tang}, {Scaife}, {Wong}, {Kapi{\'n}ska},
  {Rudnick}, {Shabala}, {Seymour}, \& {Norris}}]{tang20}
{Tang}, H., {Scaife}, A.~M.~M., {Wong}, O.~I., {et~al.} 2020, \mnras, 499, 68

\bibitem[{{Tingay} {et~al.}(2013){Tingay}, {Goeke}, {Bowman}, {Emrich}, {Ord},
  {Mitchell}, {Morales}, {Booler}, {Crosse}, {Wayth}, {Lonsdale}, {Tremblay},
  {Pallot}, {Colegate}, {Wicenec}, {Kudryavtseva}, {Arcus}, {Barnes},
  {Bernardi}, {Briggs}, {Burns}, {Bunton}, {Cappallo}, {Corey}, {Deshpande},
  {Desouza}, {Gaensler}, {Greenhill}, {Hall}, {Hazelton}, {Herne}, {Hewitt},
  {Johnston-Hollitt}, {Kaplan}, {Kasper}, {Kincaid}, {Koenig}, {Kratzenberg},
  {Lynch}, {Mckinley}, {Mcwhirter}, {Morgan}, {Oberoi}, {Pathikulangara},
  {Prabu}, {Remillard}, {Rogers}, {Roshi}, {Salah}, {Sault}, {Udaya-Shankar},
  {Schlagenhaufer}, {Srivani}, {Stevens}, {Subrahmanyan}, {Waterson},
  {Webster}, {Whitney}, {Williams}, {Williams}, \& {Wyithe}}]{mwa}
{Tingay}, S.~J., {Goeke}, R., {Bowman}, J.~D., {et~al.} 2013, \pasa, 30, e007

\bibitem[{{Tully}(2015)}]{Tully15}
{Tully}, R.~B. 2015, \aj, 149, 171

\bibitem[{{van Ojik} {et~al.}(1997){van Ojik}, {Roettgering}, {Miley}, \&
  {Hunstead}}]{vanOjik97}
{van Ojik}, R., {Roettgering}, H.~J.~A., {Miley}, G.~K., \& {Hunstead}, R.~W.
  1997, \aap, 317, 358

\bibitem[{{V{\'e}ron-Cetty} \& {V{\'e}ron}(2010)}]{veron10}
{V{\'e}ron-Cetty}, M.~P. \& {V{\'e}ron}, P. 2010, \aap, 518, A10

\bibitem[{{Willis} {et~al.}(1974){Willis}, {Strom}, \& {Wilson}}]{willis74}
{Willis}, A.~G., {Strom}, R.~G., \& {Wilson}, A.~S. 1974, \nat, 250, 625

\bibitem[{{Zovaro} {et~al.}(2022){Zovaro}, {Riseley}, {Taylor}, {Nesvadba},
  {Galvin}, {Malik}, \& {Kewley}}]{Zovaro22}
{Zovaro}, H. R.~M., {Riseley}, C.~J., {Taylor}, P., {et~al.} 2022, \mnras, 509,
  4997

\end{thebibliography}


\end{document}